# HIGH QUALITY OF SERVICE ON VIDEO STREAMING IN P2P NETWORKS USING FST-MDC

Suresh Jaganathan[1] and Jeevan Eranti[2]

[1]Department of Computer Science & Engineering,
Sri Sivasubramania Nadar College of Engineering,
Chennai, Tamil Nadu, India
whosuresh@gmail.com
[2]Department of Computer Science & Engineering,
Sri Sivasubramania Nadar College of Engineering,
Chennai, Tamil Nadu, India
jeevan4007@gmail.com

## ABSTRACT

*Video streaming applications have newly attracted a large number of participants in a distribution network. Traditional client-server based video streaming solutions sustain precious bandwidth provision rate on the server. Recently, several P2P streaming systems have been organized to provide on-demand and live video streaming services on the wireless network at reduced server cost. Peer-to-Peer (P2P) computing is a new pattern to construct disseminated network applications. Typical error control techniques are not very well matched and on the other hand error prone channels has increased greatly for video transmission e.g., over wireless networks and IP. These two facts united together provided the essential motivation for the development of a new set of techniques (error concealment) capable of dealing with transmission errors in video systems. In this paper, we propose an flexible multiple description coding method named as Flexible Spatial-Temporal (FST) which improves error resilience in the sense of frame loss possibilities over independent paths. It introduces combination of both spatial and temporal concealment technique at the receiver and to conceal the lost frames more effectively. Experimental results show that, proposed approach attains reasonable quality of video performance over P2P wireless network.*

## 1. INTRODUCTION

The demand for media services over the Internet as well as wireless networks has rapidly increased in recent years [1]. Streaming is a technology used for providing multimedia content distribution between (or amid) clients in a range of multimedia applications on the wireless network. With this knowledge, the client can playback the media substance without waiting for the complete media file to arrive. Contrast with conventional data communication, the distribution of multimedia data has severe additional problems on network bandwidth, delay and loss.

Among different methods intended for video streaming to resolve the disputes on packet-based and best-effort Internet today, distributed streaming over joint P2P network provides solution are self organizing networks with a collective large number of heterogeneous computers called nodes or peers. Each node has a certain storage space and uplink bandwidth reserved for participating as a serving node. Compared to the central server approach, the proposed system requires very low initial set-up cost and can be more scalable, reliable and each node acts both as a client and server.

Peer-to-Peer (P2P) [2] systems have increased remarkable intentions during these years. Information got among various peers is used by other peers, which runs front and back to the end users, contrast to the traditional distributed methods derived from real client/server model. Peer-to-peer video streaming systems present the similar advantages as peer to peer file transfer network but face further challenges, since data transfer needs to occur continuously to avoid play out disruptions. The characteristics of P2P system make them a better choice for multimedia content sharing or streaming over wireless network. P2P system are dynamic in nature where nodes can join and leave the network frequently and that might not have a permanent IP addresses and detect dynamic changes over the inter connection links. Virtual networks are built on the top of these networks at the application level in which individual peers communicate with each other and share both communication and storage resources, ideally directly without using a dedicated server. P2P media sharing uses two basic concepts. First, the live streaming technique broadcasts live video substances to all the peers in real time. Second, the on-demand video (VoD) streaming technique facilitates peers to enjoy the flexibility of watching a video. Recently, peer-to-peer technologies quickly attracted people and had a large success for file distribution applications as well as for other varieties of applications. A popular approach for information exchange in the P2P framework is the simultaneous streaming from multiple senders. This approach yields a higher quantity and a better tolerance to loss and delay caused by network congestions.

The rest of the paper is organized as follows. Section 2 describes the multiple description coding technique. Section 3 deals related works in MDC and EC. Section 4 presents the proposed error concealment method. Section 5 reveals the experimental results and finally, Section 6 concludes the paper.

## 2. MULTIPLE DESCRIPTION CODING

Conventional video compression standards employ a similar design which referred as single-condition systems, since they have a single state (e.g. the previous coded frame) which if lost or damaged and, can lead to the deficit or severe ruin of all consequent frames until the state is reinitialized (the prediction is refreshed).

Multiple Description Coding (MDC) [3, 4] is an approach proposed to address overcome the traffic loss over transmission channels. It is a source coding technique that generates multiple, equally important bitstreams, called descriptions, for a single video file. Different levels of reconstructed video qualities can be achieved by successfully decoding different subsets of descriptions. The advantage of doing this is that descriptions can be streamed to a receiver using disjoint streaming paths, which can potentially increase the resilience to packet loss. Unlike scalable coding, there is no interdependency among MDC descriptions, and every description can be separately decoded, successfully decoding more descriptions results in better video quality. This feature makes MDC appealing for use in the design of a concurrent video streaming system.

## 2.1. MDC Encoder

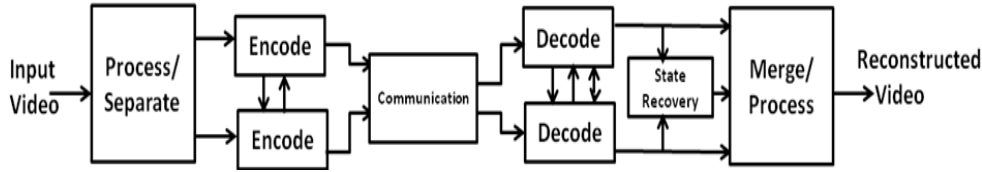

Figure 1: Multiple Description Coding

Figure 1 shows working of multiple description coding method, the original video sequence which is encoded into two sub sequences of frames (even and odd), and they are divided into two separate bitstreams. Specifically, each stream has a different state and a different prediction loop, which can be independently decoded to produce a signal of basic quality. In general there can be multiple coded streams each with its own state referred as Multiple States or Multiple State Streams as shown in Figure 2. The proposed method is conceptually related to multiple description coding, however it differs in the account used for each description and most importantly in its use of state recovery.

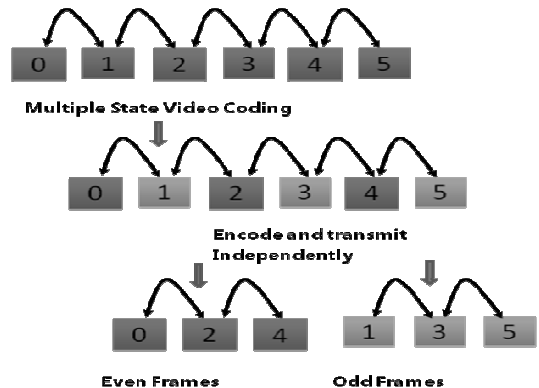

Figure 2: Multi State Coding

The different streams should be transmitted over different channels undergoing independent error effects to minimize the chance that both streams are lost. For example, the bitstreams from the even and odd frames can be sent in different packets over a packet network, so that any lost packet will only affect one of the streams.

## 2.2. MDC Decoder

Similar to the encoder, the decoder alternates the previous decoded frame to perform the prediction. If there are no errors and both the even and odd streams are received correctly, then both streams are decoded to produce the even and odd frames which are interleaved to produce basic video display.

If a stream has an error then the state for that stream is incorrect and there is a error propagation for that stream. However, the other independently decodable stream can still be accurately and straightforwardly decoded to produce usable video. For example, if the bitstream corresponding to the odd frames is lost, the even frames may still be decoded and displayed, recovering the video at half its original frame rate. The novelty in the multiple state coding processes is that it provides improved error concealment and enables state recovery of the lost stream. In case of error concealment that provides access both previous and future frames can greatly assist in recovering the corrupted stream or frame and thereby restore the video to its full frame rate.

## 3. RELATED WORKS ON MDC AND EC

MDC is particularly beneficial for delay-sensitive, real-time applications such as streaming video, in which packet losses may significantly corrupt the quality. Conventional approaches to combat channel errors such as Automatic Retransmission reQuest (ARQ) and Forward Error Correction (FEC) require retransmission of the lost packets or to addition of redundant bits for the purpose of error detection and correction. However, ARQ-based approaches are not applicable in applications when a back-channel is not available or when the retransmission delay is not acceptable. For FEC-based approaches, because of the highly varying network conditions, it is difficult to adaptively adjust the amount of redundancy, which makes the FEC either inefficient or ineffective.

A simple and practical MDC scheme known as Multiple Description Scalar Quantizer (MDSQ) was proposed by Vaishampayan et al. [5]. In this scheme, two sub-streams are generated by producing two indices for each quantization level. The index assignment is designed to be equivalent to a fine quantizer when both indices are received, but a coarse quantizer when only one index is received. One simple implementation can be created by using two quantizer whose decision regions are offset by half a quantization step size. Another MDC scheme available is Multiple Description Transform Coding (MDTC) [6]. In this ideal source coding, the coefficients used to represent the signal are uncorrelated as possible to maximize the coding efficiency. However, under this paradigm it is very difficult to estimate the value of a lost coefficient from those that remain. To achieve robustness against coefficient losses, the transform coefficients can be divided into multiple groups where the correlations within each group are minimized while inter-group correlations are tolerated.

The Recursive Optimal per-Pixel Estimate (ROPE) [7] algorithm allows the encoder to estimate the pixel-by-pixel predictable distortion of the decoded video appropriate to channel failure. This algorithm needs an input regarding approximation of the packet deficit rate and the information on error concealment method employed in the decoder. An extended ROPE algorithm accurately estimates the rate distortion due to various loss patterns and applies it for optimal mode selection using rate-distortion optimization. The rate distortion selection scheme causes a slight performance degradation while providing advantages of finer priorities over network transmission and lower complexity. Finally, it exhibits high computational costs and long encoding time.

A Slepian-Wolf based inter frame transcoding (SWIFT) scheme [8] transcodes an inter-coded P-blocks to a new type of block called X-block for the purpose of error robustness. The X-block can be losslessly transcoded back to the exact original P-block when there is no transmission error, and can also be decoded robustly like an I-block when there are errors in the predicted block. The compression of the proposed X-block is based on Slepian Wolf coding (SWC) which can achieve partially intra-like robustness with inter-like bit rate. SWIFT method improves the error resilience capability of off-line compressed video. At the decoder, the transcoded video can be converted back to the original compressed video in error free case and can also be robustly decoded when error occurs.

Historically, MDC and EC algorithms have evolved separately. For MDC, most research efforts focus on generation of multiple descriptions with as little redundancy as possible, but still with the capability that each description provides an acceptable quality signal. In a typical multiple description codec, the decoder simply reproduces the signal using the received description(s). Error concealment methods, on the other hand, have dealt mainly with the loss of entire Macro Blocks (MB) or DCT coefficients. However, we can envision using a similar technique to recover the lost descriptions when an MDC system is employed.

We propose an algorithm that exploits the spatial/temporal correlation to help reconstruct the video signal in the presence of lost descriptions. If both descriptions are received, the decoder simply reconstructs the signal from two descriptions. If only one description is received, the decoder is designed to reconstruct the signal by maximizing a spatio-temporal smoothness measure within the constraints of the received description.

# 4. PROPOSED SYSTEM

Streaming compressed video over communication channels present losses, errors and extreme delay, these channels need countermeasures to preserve the quality of the viewing occurrence. There are two techniques to overcome the errors and delay: error resilience and error concealment.

## 4.1. Error Resilience

Error control technique help mitigate the impact of transmission errors or packet loss on the quality of the decoded video. They are essential for applications which are difficult to achieve timely and error-free delivery of the stream. Examples include, interactive applications requiring low latencies, i.e. two-way video communications, Internet-based television, etc., or situations where the network fabric is unreliable, as in case of P2P application or wireless links. Error Resilient Encoding is a normal method for transmission errors which should not lead to unacceptable distortion in the reconstructed video [9]. The main objective is maximum gain in error resilience with minimum redundancy. It provides 3 basic techniques: i) better error detection ii) concealment at the decoder, to prevent error propagation, and iii) guarantee basic level of quality.

Error resilience, in the context of video coding systems can be extrapolated as a system to withstand errors and perform acceptably, thus providing a good video quality at the receiver end. Error resilience functionality in the encoder produces a bitstream that supports error recovery at the decoder.

## 4.2. Error Concealment

Error concealment is an effective technique since no overhead or redundancy is required [10, 11, 12]. Error Concealment in the context of video systems can be extrapolated to be the action of hiding the effects of errors. After the errors are detected and as much as probable specifically localized, it is time to initiate concealment, which basically means to conceal their negative subjective effects. The decoder tries to minimize the negative impact of the errors in the decoded images just by using the available decoded data (motion, texture or also shape) or in some conditions additional data may be requested from the encoder. The concealment process is always based on some sensible assumptions, which may depend on the type of application and content involved.

### 4.2.1. Spatial Error Concealment

In this type of concealment, only data from the current time instant is used to perform the concealment operation. This basically means that corrupted areas in a given image are recovered by interpolating the data from the surrounding correctly decoded area of the same image.

As it is easy to understand, this approach can have serious problems if the corrupted area of the image that is being concealed represents quite a large part of the whole image, especially if it is very heterogeneous. This is the reason why these techniques are mostly used for large images, where the corrupted areas are typically small when compared to the size of the whole image. In particular, this technique works well for homogenous areas. In addition, these techniques are also suggested for sequences where there is very little temporal redundancy. Finally, spatial error concealment attempts to use the data in the present frame to restore the lost data.

### 4.2.2. Temporal Error Concealment

In this case, the data of other time instants is used to perform the concealment. The most common approach is to use the data of the previous time instant. When this technique is applied to sequences where a large temporal [13, 14] redundancy exists, which corresponds to most of the cases, the results will typically be very good. Otherwise, serious problems can appear if the decoder tries to conceal the image in the current time instant with data from images in surrounding time instants that have very little to do with it, for example in a scene cut.

## 4.3. Hybrid Error Concealment

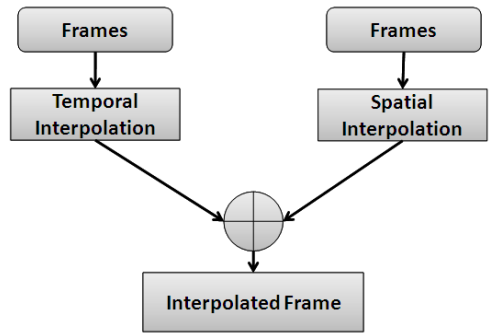

Figure 3: Hybrid Error Concealment

Figure 3 shows the combination of the previous two, in the sense that data from both the current time instant and other time instants is used. These techniques estimate motion and create an image by displacing objects (e.g., pixels or macro blocks).The idea behind the spatial-temporal concealment technique is to have the best of both worlds by combining the spatial and the temporal concealment techniques [15,16,17]. This way, some parts of the image might be concealed using spatial concealment, others by using temporal concealment and still others by using a little bit of both. Ideally, each of these solutions should be used for the parts of the image that are spatially, temporally or both.

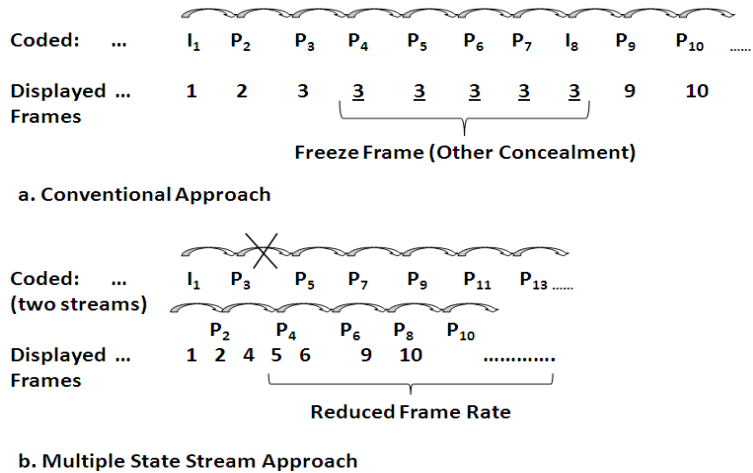

Figure 4: Multiple State Stream Approach

Figure 4 depicts multiple state stream approach. There is an error in decoding of frame that depends on the frame P3. In a conventional single-state approach (Figure 4a) frame 4 is lost and the decoder may freeze frame 3 (or performs other error concealment) until the next I-frame. In a simple two-state stream approach (Figure 4b) coded stream #1 is lost, however displayed stream #2 can be accurately decoded- recovering the video at half its original frame rate but without any other distortions. More importantly, coded stream #1 can often be recovered by appropriately using displayed stream #2.

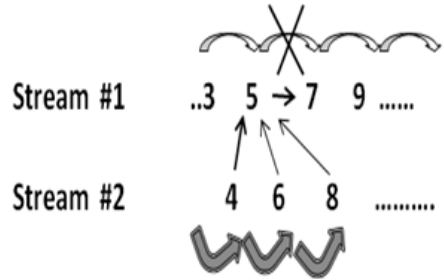

Figure 5: Concealment Approach

Figure 5 shows the correctly received streams enable improved error concealment and potential state recovery for the lost stream. Frame 5 (Stream #1) may be concealed or recovered by using information from previous (temporal) and future (spatial) as shown in Stream #2 and decoded correctly using frames signified by the dashed lines.

## 5. EXPERIMENTAL RESULTS

The performance of the proposed method is tested on several test sequences such as Akiyo, Foreman, Lena and Sachin of image size *352x258*. The proposed scheme Flexible spatial temporal concealment (FST) achieves better performance than the conventional scheme only with spatial concealment. Table 1 shows the results of these comparisons in terms of PSNR (dB).

Table 1: PSNR values (in dB) of test images reconstructed by two algorithms

| Image Error Ratio | Akiyo | | Foreman | | Lena | | Sachin | |
|---|---|---|---|---|---|---|---|---|
| | Spatial | FST | Spatial | FST | Spatial | FST | Spatial | FST |
| 10% | 24.68 | 26.81 | 27.73 | 29.32 | 23.32 | 25.37 | 28.52 | 32.43 |
| 20% | 23.32 | 25.32 | 26.19 | 28.79 | 22.32 | 24.69 | 26.29 | 30.97 |
| 30% | 22.29 | 24.32 | 25.44 | 27.01 | 21.32 | 23.40 | 24.81 | 28.75 |
| 40% | 21.10 | 23.96 | 24.26 | 26.04 | 20.32 | 22.88 | 22.32 | 26.21 |
| 50% | 20.20 | 22.80 | 23.17 | 25.12 | 19.90 | 21.06 | 20.90 | 24.06 |

Table 1 show the results got for various sequences against error ratio. The sequence Akiyo has lesser motion when compared to Foreman and Sachin. Lena sequence is a nearly a still image and has very less motion. From the table it can be seen that when the error ratio is less than or equal to 10%, there is an increase in PSNR value for our proposed method (26.81) when compared to only spatial concealment method (24.68). For Sachin sequence, which has more motion, the PSNR value obtained is 32.43 for 10% error ratio and for 50% error ratio it is 24.06, which clearly indicates that proposed method with hybrid concealment recovers the image with a visible quality.

Figure 6 show the graph plotted using the values from Table 1. From the graph, it is clear that the proposed method significantly improves the PSNR performance compared with other schemes. For the "Akiyo", "Foreman" and "Lena" images, a PSNR gain of about 2 dB can be achieved using the proposed algorithm. For the "Sachin" image, the improvement is even greater.

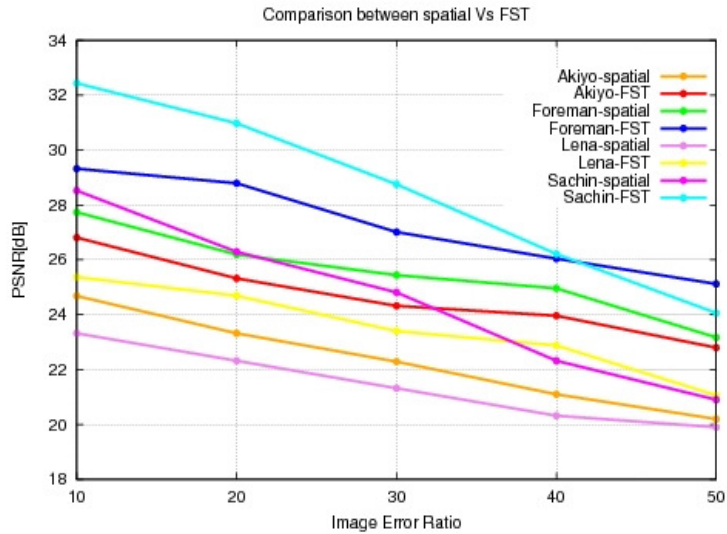

Figure 6: PSNR performances in terms of the Image Error Ratio

Figure 7 show the subjective comparison results of two different concealment algorithms. One is the spatial concealment algorithm and the other is proposed FST concealment algorithm.

From both figures we can observe that the spatial concealment algorithm (Fig. 7 (c)) restores the missing areas, but at the expense of significant blurring effects around the edges of the recovered areas. In Fig. 7 (d) however, with the proposed scheme, it can be observed that many dominant edges have been reconstructed gracefully.

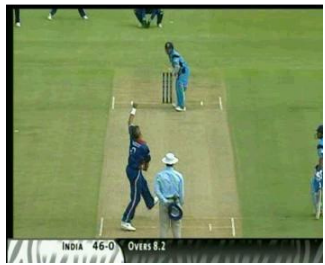 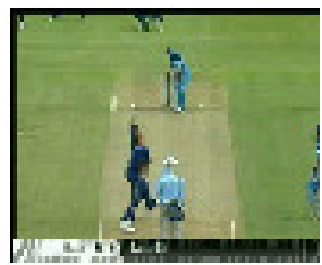

7a. Original Image          7b. Damaged Image

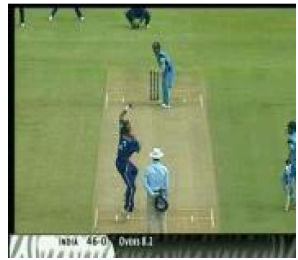

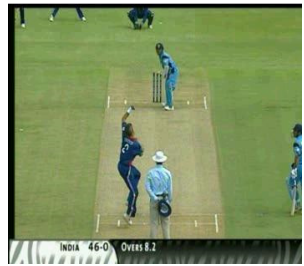

7c. Reconstructed Image using Spatial Concealment Method

7d. Reconstructed Image using FST Method

Figure 7: Subject Comparison of different algorithms for the image "Sachin"

| | Akiyo | Foreman | Lena | Sachin |
|---|---|---|---|---|
| Original Image | 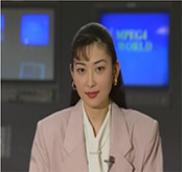 | 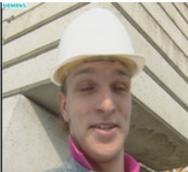 | 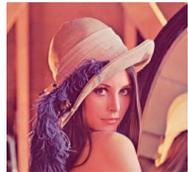 | 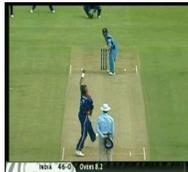 |
| Corrupted Image | 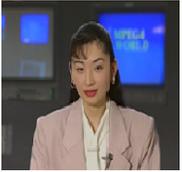 | 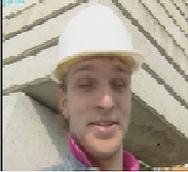 | 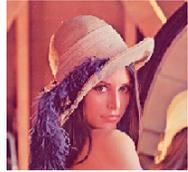 | 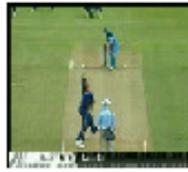 |
| Reconstructed Image | 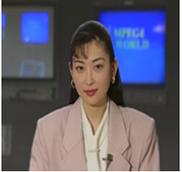 | 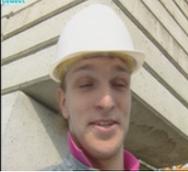 | 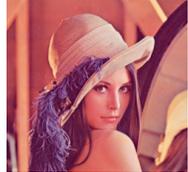 | 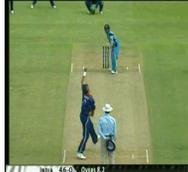 |
| *PSNR(dB) (Proposed)* | *22.80* | *25.12* | *21.06* | *24.06* |

Figure 8: PSNR comparisons of the video sequences (a) Akiyo (b) Foreman (c) Lena (d) Sachin with respect to difference packet loss rates.

Figure 8 shows experimental results and presents the PSNR for all four different sequences, such as Akiyo, Foreman, Lena and Sachin. The results show that the PSNR for the frames after the lost or corrupted frame can be quickly resumed and the proposed algorithm is obviously more efficient in stopping error propagation than other methods. In summary, MDC with FST presents a promising approach for error resilient video streaming over the error prone networks.

## 6. CONCLUSION

In this paper we have proposed a new error concealment method called FST-MDC a combined with MDC a promising approach for error concealment in video streaming over error prone networks. Our approach adopts a combination of spatial and temporal, which takes privileges in data correlation of these two domains for better estimation of lost descriptions or damaged descriptions. At the receiver end, this approach can recover the damaged images itself without adding the extra information. Experimental results show that the proposed method efficiently recovers the detailed content and the PSNR quality is improved about 2 ∼ 5 dB with respect to the conventional spatial concealment algorithms. When the image error ratio goes more than 50%, then the image is mostly corrupt and our approach conceals to some extent with acceptable viewing quality. Presently FST approach checks and conceals a single frame at a time, when multi-frames are lost, modification has to be done in FST approach which is an ongoing work.

## Authors

**Suresh Jaganathan** received his Bachelors Degree from Mepco Schlenk Engineering College, Sivakasi and M.E (Software Engg.) from Anna University, Chennai. Currently he is pursuing his PhD in Jawaharlal Nehru Technological University (JNTU), Hyderabad, in the field of Grid Computing. To his account he has published 13 papers in the area of, Grid Computing and Peer-to-Peer Networks in reputed National & International Conferences. He has 15 years of experience in Teaching and is currently working as an Assistant Professor in Department of Computer Science & Engineering in Sri Sivasubramania Nadar College of Engineering, Chennai. His research interests are Grid Computing, Distributed Computing and Neural Networks. He is a Member of IEEE and Life Member of CSI and ISTE in India.

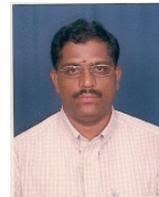

**Jeevan Eranti** received his Bachelors Degree from Kodaikanal Institute of Technology, Kodaikanal and M.E (CSE) from Sri Sivasubramania Nadar College of Engineering, Chennai, India. Shortly he is going to join as Project Engineer in Wipro Technologies Ltd, Bangalore, India. His area of interest includes Multimedia Communications, Peer-to-Peer Networks and Image Processing.

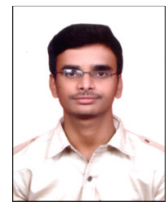